\documentclass{svmult}
\usepackage{graphicx}
\usepackage{latexsym}
\usepackage{subfigure}

\title{A game theoretical perspective on the somatic evolution of cancer}

\author{David Basanta*, Andreas Deutsch}
\institute{Technische Universit\"{a}t Dresden, Zentrum f\"{u}r Informationsdienste und Hochleistungsrechnen, N\"{o}thnitzer-str. 46, 01187, Dresden, Germany\\email:david.basanta@moffitt.org}

\makeindex 
\date{ }
\begin{document}
\maketitle


\section{Introduction} 

Environmental and genetic mutations can transform the cells in a co-operating healthy tissue into an ecosystem of individualistic tumour cells that compete for space and resources \cite{nowell:1976,Crespi:2005,Merlo:2006}. If we consider a tumour as an ecosystem it is possible to utilise tools traditionally used by ecologists  to study the evolution of a population in which there is some degree of phenotypical diversity. One such tool is evolutionary game theory (EGT) which merges traditional game theory with population biology \cite{Sigmund:1999}. It allows the prediction of successful phenotypes and their adaptation to environmental selection forces. EGT is considered as a promising tool in which to frame oncological problems  \cite{gatenby:2003d} and has  been recently made more relevant by phenotypic studies of carcinogenesis such as the ones by Hanahan, Weinberg and colleagues  \cite{hanahan:2000,Hahn:2002}.

Game theory (GT) was introduced by von Neumann and Morgenstern as an instrument to study human behaviour \cite{Neumann:1953,Nowak:2006}. A game describes the interactions of two or more players that follow two or more well defined strategies in which the benefit of each player (payoff) results from these interactions \cite{Merston-Gibbons:2000}. GT can be employed to study situations in which several players make decisions in order to  maximise their own benefit. GT was initially introduced to model problems in economics, social and behavioural sciences and is used as a formal way to analyse interactions between agents that behave strategically. 
Evolutionary game theory is the application of conventional GT as used by economists and sociologists to study evolution and population ecology \cite{Sigmund:1999}.  As opposed to conventional GT, in EGT the behaviour of the players is not assumed to be based on rational payoff maximisation but  it is thought to have been shaped by trial and error - adaptation through natural selection or individual learning \cite{Maynard:1982}. In the context of the evolution of populations there are two GT concepts that have to be interpreted in a different light. First, a strategy is not a deliberate course of action but a phenotypic trait. The payoff is Darwinian fitness, that is, average reproductive success. Secondly, the players are members of a population that compete or cooperate to obtain a larger share of the population \cite{Sigmund:1999}. 

To illustrate some of the ideas in EGT let us consider the following example named the Hawk-Dove game \cite{Maynard:1982}. In this game we study an imaginary population of individuals and a resource V which affects the reproductive success of the individuals in this population. The population contains two phenotypes that represent two different strategies to access the resource. When two individuals compete for the resource the outcome will depend on the phenotypic strategies involved. The first phenotype, called Hawk in the game, always escalates the fight until injured (at a cost in fitness equal to C) or  until the rival retreats. The second phenotype, known as Dove in the game, will retreat if the opponent escalates, that is, if the opponent seems determined to fight. The interactions between the different phenotypes are shown in the payoff table \ref{tab:HawkDove}.

\begin{table}[h]        
	\begin{center}
	\caption{Payoff table for the change in fitness in the Hawk-Dove game.}
	\begin{tabular}{|c|c|c|} \hline
	& Hawk                &      Dove     \\ \hline
	Hawk           &       $\frac{V-C}{2}$  &    0      \\ \hline       
	Dove            & V  & $\frac{V}{2}$       \\ \hline
	\end{tabular}  
	\label{tab:HawkDove}
	\end{center}
\end{table}

Table \ref{tab:HawkDove} presents the interactions between the different phenotypes considered in the game. The table should be read following the columns such that the payoff for a Hawk playing another Hawk is $\frac{V-C}{2}$ expressing the fact that they both have to share the resource and that they stand an equal chance of getting injured. The payoff of a Hawk playing a Dove is V since the Dove will withdraw from the competition. A Dove playing a Hawk gets no payoff since it withdraws and when playing another Dove it will get the resource V in half of the occasions. With this information it is possible to predict that if the population is mainly composed of individuals with the Dove phenotype then a Hawk individual will have a significant fitness advantage (as in most of the interactions, the rivals are likely to be a  Dove and thus retreat from a full scale fight for the resource). On the other hand if the fitness cost of injury is more than twice as high as the benefit provided by the contested resource then a Dove would be quite successful in a population dominated by Hawks (since a Hawk that interacts frequently with other Hawks is likely to be eventually wounded and a Dove will always avoid costly wounds). Using this example Maynard Smith introduced the concept of an evolutionary stable strategy (ESS) \cite{Maynard:1982}. An ESS is defined as a phenotype that, if adopted by the vast majority of a population,  will not be displaced by any other phenotype that could appear in the population as a result of evolution \cite{Maynard:1982}. Under this definition the Dove phenotype cannot be an ESS and only under some specific circumstances (when the fitness benefit of getting the resource outweighs the fitness cost of an injury) would a Hawk phenotype be evolutionary stable. 

\begin{figure}
	\centering
		 \includegraphics[scale=0.7]{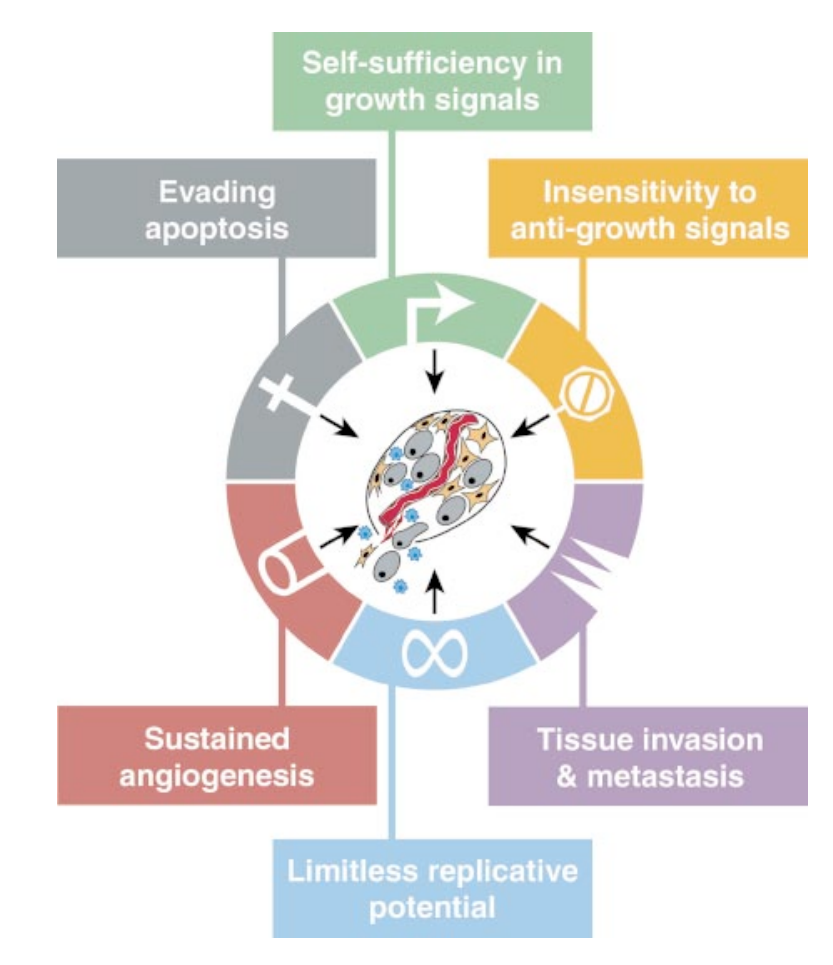}
	\caption{Acquired capabilities of cancer: Hanahan and Weinberg suggest that most if not all cancers have acquired the same set of capabilities during their evolution. These capabilities are:  unlimited replicative potential, environmental independence from anti growth signals, production of own growth signals, evasion of apoptosis, angiogenesis and invasion (from \cite{hanahan:2000} with permission).}
	\label{fig:hanahan}
\end{figure}

GT has been used to address many problems in biology in which different species or phenotypes within one species compete. Examples of this are  the evolution of sex ratios \cite{Fisher:1930}, the emergence of animal communication \cite{Smith:2003} and fighting behaviour and territoriality \cite{Maynard:1982}. A recent focus on the capabilities that cells acquire as tumours evolve \cite{hanahan:2000} has shown how the interplay between different phenotypes with different capabilities can lead to different evolutionary paths. This stresses the importance of GT as a modelling tool in cancer. The most important capabilities which cells have to acquire in a neoplasm that will become a malignant tumour are shown in figure \ref{fig:hanahan}. They include: unlimited replicative potential, environmental independence for growth, evasion of apoptosis, angiogenesis and invasion. The circumstances in which these capabilities evolve and spread through the tumour population can be studied using GT. Some of the most important milestones in the transformation of a healthy tissue cells into malignant cancer such as tumourigenesis, angiogenesis and invasion, have already been approached with GT.

The remaining of this chapter will provide, to the best of our knowledge, all the relevant examples of the application of GT to the study of the somatic cancer evolution and finally hint some of the possible future venues of this method in the context of cancer research.

\section{Tumourigenesis}
Tumour initiation requires the acquisition of a number of phenotypic capabilities such as evasion of apoptosis and independence from environmental signals (see  figure \ref{fig:hanahan}). The evolution of these capabilities, normally acquired when the tumour is still in the avascular stage, are studied in research by Tomlinson and Bodmer \cite{tomlinson:1997b,tomlinson:1997a} and by Gatenby and Vincent \cite{gatenby:2003a}.

\subsection{Evasion of apoptosis}

\textbf {Problem}. Apoptosis or programmed cell death is a mechanism that hinders tumour progression. Cells with a working apoptotic machinery die when genetic abnormalities are detected \cite{hanahan:2000}. Thus cells in a malignant cancer have to evolve mechanisms to disable the apoptotic machinery. 

\textbf {Model}. Tomlinson and Bodmer \cite{tomlinson:1997b} present a model in which three different apoptosis evasion related strategies are considered:
\begin{enumerate}
\item Cells that produce a paracrine growth factor to prevent apoptosis of neighbouring cells.
\item Cells that produce an autocrine growth factor to prevent apoptosis of themselves.
\item Cells susceptible to paracrine growth factors but incapable of production of factors.
\end{enumerate}

The aim of the model is to study the possibility of stable coexistence of the different phenotypes (polymorphism) that could be possible in a tumour when only these three phenotypes are considered. 

\begin{table}[h]        
	\begin{center}
	\caption{Payoff table for the game of programmed cell death. Parameter a represents the cost of producing a paracrine factor, b is the cost of the factor produced in an autocrine fashion, c is the fitness benefit of evading apoptosis.}
	\begin{tabular}{|c|c|c|c|} \hline
	& 1                &      2 & 3    \\ \hline
	1           &      $1-a+b$  &    $1+b+c$ & $1+b$       \\ \hline       
	2           &       $1-a$ &    $1+c$   & 1    \\ \hline       
	3             & $1-a$  & $1+c$    & 1   \\ \hline
	\end{tabular}  
	\label{tab:apoptosisTB}
	\end{center}
\end{table}

In table \ref{tab:apoptosisTB}, \textit{a} is the cost of producing the paracrine factor, \textit {b} the benefit of receiving the paracrine factor and \textit {c} the benefit of producing the autocrine factor. 

\textbf {Results}. If \textit {a} is positive then the third strategy displaces the first one from the population. If only the other two strategies are considered then if the benefit of the autocrine factor, provided by \textit {c}, is positive the second strategy will displace the third one. In most relevant situations the model shows a strong selection for the autocrine factor producing phenotype and under the assumptions of the model the altruistic strategy (the first one) will always be displaced.

\textbf {Remarks}. The model is very simple and easy to understand while at the same time it captures the relevant features necessary to study the evolution of the mechanisms to avoid apoptosis. However the authors do not explain the mechanisms by which these phenotypes could appear in a tumour and what would be the biological explanation of the costs and benefits of the different growth factors.

\subsection{Environmental poisoning}

\textbf {Problem}. Tomlison introduced a further model in which he considers the hypothesis that tumour cells might boost their own replicative potential at the expense of other tumour cells by evolving the capability of producing cytotoxic substances \cite{tomlinson:1997a}. 

\textbf {Model}. Tomlinson speculates with different strategies that cells may adopt to produce or cope with toxic factors. The main model aims to study the polymorphic equilibria when cells can adopt one of the three following strategies:

\begin {enumerate}
\item Cells producing cytotoxic substances against other cells,
\item cells producing resistance to external cytotoxic substances, and
\item cells producing neither cytotoxins nor resistance.
\end{enumerate}

Table \ref{tab:toxin1T} shows the payoff table of the game with these phenotypes.

\begin{table}[h]        
	\begin{center}
	\caption{Payoff table for the change in fitness for cells in a tumour in which the base payoff is z, e the cost of producing the cytotoxin, f the fitness cost of being affected by the cytotoxin, g the advantage of subjecting another cell to the cytotoin and finally the cost of developing resistance to the cytotoxin is h.}
	\begin{tabular}{|c|c|c|c|} \hline
	& 1                &      2 & 3    \\ \hline
	1           &      z-e-f+g  &    z-h & z-f       \\ \hline       
	2           &       z-e &    z-h   & z    \\ \hline       
	3             & z-e+g  & z-h   & z   \\ \hline
	\end{tabular}  
	\label{tab:toxin1T}
	\end{center}
\end{table}

\textbf {Results}. Game theoretical analysis and simulations show that production of cytotoxic substances against other tumour cells can evolve in a tumour population and that several cytotoxin related strategies may be present at a given time (polymorphism). 

\textbf {Remarks}. Although the author admits that there is little experimental evidence for mutations that cause tumour cells to harm their neighbours, the Warburg effect could fit nicely in the framework presented in this work. The Warburg effect describes the switch of tumour cells from the conventional aerobic metabolism to the glycolytic metabolism. This metabolism is less efficient but produces, as a by-product, acid that can harm neighbouring cells \cite{warburg:1930,gatenby:2003c,gatenby:2004,Gatenby:2006}. Thus, in the game described by Tomlinson, e could correspond to the fitness loss of the less efficient glycolytic metabolism, f could be the fitness loss of a normal cell in an acid environment and g the fitness benefit received by glycolytic cells that can take advantage of the harm done to their non glycolytic neighbours.

\section{Angiogenesis}

\textbf {Problem}. A very important capability that has to be acquired by tumours on the path to cancer is angiogenesis. Without access to the circulatory system tumours do not grow to sizes bigger than 2mm in diameter \cite{Folkman:1992}. Cells capable of angiogenesis produce growth factors that promote the creation of new blood vessels that can provide nutrients and oxygen to previously unreachable areas in a growing tumour. Presumably the factors will be produced at a cost to the tumour cell. 

\textbf {Model I}. In their interpretation of an angiogenic game Tomlinson and Bodmer  \cite{tomlinson:1997b} consider two strategies: cells denoted as A+ can produce angiogenic factors at a fitness cost \textit{i} and cells denoted as A- that produce no angiogenic factors. In any case cells will get a benefit \textit{j} when there is an interaction involving an angiogenic factor producing cell. The payoffs for the interactions between these cells are shown in table \ref{tab:angiogenesisTB}.

\begin{table}[h]        
	\begin{center}
	\caption{Payoff table for the change in fitness for a cell in a tumour with cells capable of producing angiogenic factors (A+) and cells susceptible to benefit from growth factors (A-).}
	\begin{tabular}{|c|c|c|} \hline
	& A+                &      A-     \\ \hline
	A+           &       1-i+j  &    1+j       \\ \hline       
	A-             & 1-i +j  & 1       \\ \hline
	\end{tabular}  
	\label{tab:angiogenesisTB}
	\end{center}
\end{table}

\textbf {Results}. The model shows that as long as the benefit j of angiogenesis is greater than the cost i of producing angiogenic factors then both types of strategies will be present in a tumour in proportion to these costs. 

\textbf {Remarks}. This model of angiogenesis is rather simplistic but constitutes a nice foundation for later models that take into account spatial considerations. A more significant drawback is that the model does not attempt to suggest a link between the different fitness costs and benefits and the underlying biological mechanisms. 

\textbf {Model II}. The first extension to this model was proposed by Bach et al \cite{bach:2001} suggesting a game with interactions between three players. In this game the benefit \textit{j} of angiogenesis is obtained only when at least two of the three players produce the angiogenic factor. The new payoff table is shown in table \ref{tab:angiogenesisThr}.

\begin{table}[h]        
	\begin{center}
	\caption{Payoff table for the change in fitness for a cell in a tumour with cells capable of producing angiogenic factors (A+) and cells susceptible to benefit from growth factors (A-). In this version the interactions involve three players. The payoff of a player is given by the columns (ie. the payoff of a A+ cell interacting with a A+ and a A- cell is 1 - i + j but a A+ interacting with two A- cells is 1 - j).}
	\begin{tabular}{|c|c|c|} \hline
	& A+                &      A-     \\ \hline
	A+, A+           &       1-i+j  &    1+j       \\ \hline       
	A+, A-             & 1-i +j  & 1       \\ \hline
	A-, A-             & 1-i   & 1       \\ \hline
	\end{tabular}  
	\label{tab:angiogenesisThr}
	\end{center}
\end{table}

\begin{figure}
	\centering
		 \includegraphics[scale=0.8]{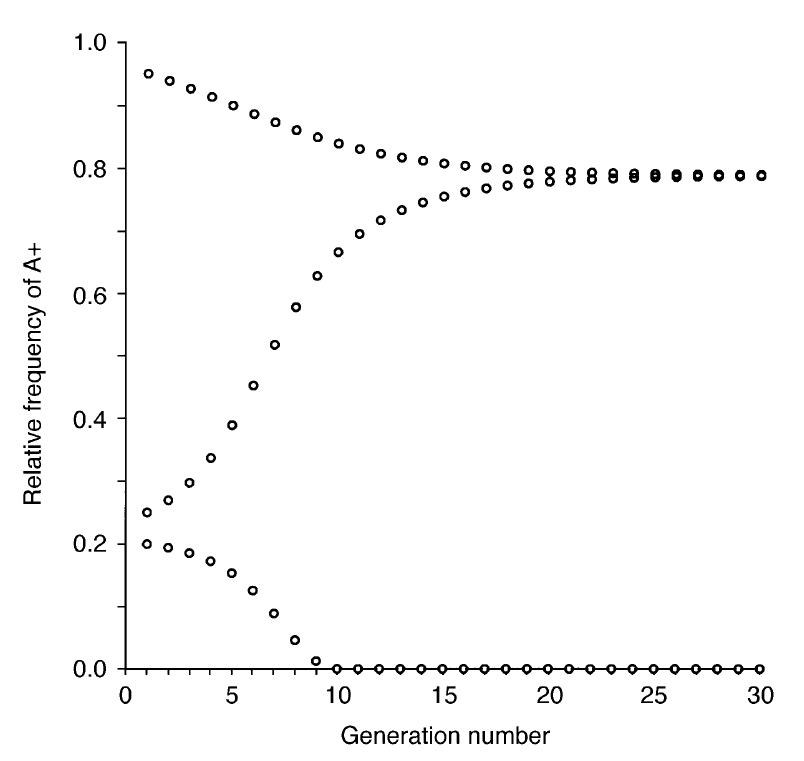}
	\caption{Simulations of the threshold model in which the benefit of angiogenesis (j) is three times the cost of producing the growth factors (i). The plot shows the results of three simulations with a different of A+/A- players in the starting population. The ordinate is the frequency of A+ cells and the abscissa is the number of cell generations. (from \cite{bach:2001} with permission )}
	\label{fig:bach3p}
\end{figure}

\textbf {Results}. The authors produce simulations using table \ref{tab:angiogenesisThr} and show (see figure \ref{fig:bach3p}) that this game yields results than differ from the original ones introduced by Tomlinson and Bodmer. In this case even when the cost i is smaller than the benefit j there are many scenarios for which the angiogenic strategy will be displaced from the population. The existence of a polymorphic equilibrium containing the angiogenic strategy depends, intuitively, on the cost of producing angiogenic factors in comparison to the benefit they give but also on the relative frequency of cells playing the angiogenic strategy in the population. These results suggest that a gene therapy against the reparation of tumour supressor genes would only need to change a fraction of the mutated cells before the dynamics of the system drives them to extinction.

\begin{figure}
	\centering
		 \includegraphics[scale=0.8]{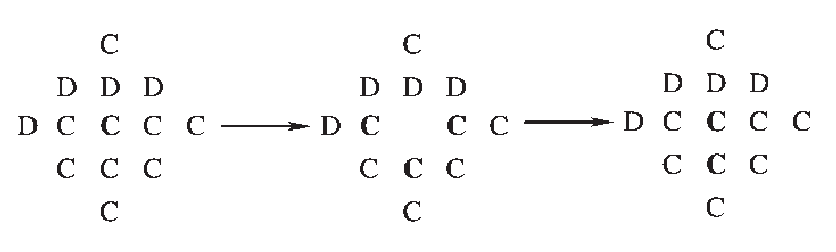}
	\caption{A diagrammatic representation of the cell update process in the spatial model of the angiogenesis game by Bach and colleagues \cite{bach:2003}. First a cell is selected to be removed (in this case the C marked in bold in the configuration on the left). Then the four neighbouring cells all assess their fitness (middle configuration). The cell with the highest fitness then reproduces into the empty space (right). (from \cite{bach:2003} with permission)}
	\label{fig:bachspa}
\end{figure}

\textbf {Model III}. In a separate research Bach et al \cite{bach:2003} studied how a spatial version of the angiogenesis game could produce different results to those of the original model by Tomlinson and Bodmer. In this game players inhabit a 100x100 lattice. Each player can follow either the angiogenic (A+) or the non-angiogenic (A-) strategy. Time is discrete and in each time step a number of cells is removed from the lattice at random. Neighbouring cells compete (using table \ref{tab:angiogenesisTB}) to occupy unallocated space. The candidate cell that achieves the highest score interacting with the neighbours determines the strategy that will be followed by the new cell in the vacant slot (see figure \ref{fig:bachspa}). 

\textbf{Results}. The authors found the results of the new formulation of the model markedly different from those of the non-spatial counterpart. In spatial models, space tends to favour growth promoters in ways that cannot be seen in the non-spatial model. In any case polymorphic equilibria do exist and in both spatial and non spatial cases the proportion of angiogenic players increases as the benefits of angiogenesis increase or as the cost of producing the growth factors decreases. The authors also conclude that, contrary to other evolutionary models  \cite{Nowak:1992}, space does not significantly favour co-operative strategies in populations of cells. 

\section{Motility/invasion}

A tumour in which cells develop the capability of invading other tissues becomes a malignant tumour and thus significantly worsens the prognosis of a patient. EGT is a tool that could greatly improve our understanding of the circumstances that influence the successful evolution of invasive phenotypes. 

\textbf {Model I}. Mansury and colleagues have recently introduced a rather unconventional EGT formulation that they used on top of a previous cellular automaton model in the context of brain tumours \cite{Mansury:2003,mansury:2006}. This GT module allows the original model to deal with cell-cell interactions in a tumour consisting of cells with different phenotypes. The model was designed to investigate the genotype to phenotype link in a polymorphic tumour cell population. This model covers two distinct strategies. Strategy P is characterised by a highly proliferative genotype and a high number of gap junctions. Strategy M is characterised by a highly migratory phenotype and a low number of gap junctions (which are used for cell to cell communication). As opposed to other game theoretical models in which a payoff table determines the fitness change of players when they interact, in this model three tables are used not to compute fitness change but to encode the rules of the cellular automata. The tables are shown in tables \ref{tab:motprolifCom}, \ref{tab:motprolifGrowth} and \ref{tab:motprolifMot}.

\begin{table}[h]        
	\begin{center}
	\caption{Payoff table that describes the rate of change in cell to cell communication skills depending on the phenotype of the interacting cell in the model by Mansury et al. \cite{mansury:2006}. }
	\begin{tabular}{|c|c|c|} \hline
	& P                &      M     \\ \hline
	P           &       $\uparrow\uparrow\uparrow$  &   $\uparrow\uparrow$       \\ \hline       
	M             & $\uparrow\uparrow$  & $\uparrow$       \\ \hline
	\end{tabular}  
	\label{tab:motprolifCom}
	\end{center}
\end{table}

\begin{table}[h]        
	\begin{center}
	\caption{Payoff table that describes how the proliferation capability of a cell is influenced by its interaction with cells with other phenotypes (proliferative or motile) in the model by Mansury et al. \cite{mansury:2006}.}
	\begin{tabular}{|c|c|c|} \hline
	& P                &      M     \\ \hline
	P           &       $\downarrow \downarrow \downarrow$  &    $\downarrow \downarrow$       \\ \hline       
	M             & $\downarrow \downarrow$  & $\downarrow$       \\ \hline
	\end{tabular}  
	\label{tab:motprolifGrowth}
	\end{center}
\end{table}

\begin{table}[h]        
	\begin{center}
	\caption{Payoff table that describes how the motility of a cell with a given phenotype changes depending on what other cells the cell is interacting with. Model by Mansury et al. \cite{mansury:2006}.}
	\begin{tabular}{|c|c|c|} \hline
	& P                &      M     \\ \hline
	P           &       $\uparrow$  &   $\uparrow\uparrow$       \\ \hline       
	M             & $\uparrow\uparrow$  & $\uparrow\uparrow\uparrow$       \\ \hline
	\end{tabular}  
	\label{tab:motprolifMot}
	\end{center}
\end{table}

Table \ref{tab:motprolifCom} shows how the cell to cell communication capabilities change according to the strategies followed by the interacting cells. Since the authors assume that there is a negative correlation between the extent of communication and proliferation then the proliferative capability of a cell with the proliferative strategy P will be lower if it interacts with another cell with the same strategy. On the other hand it will be higher if it interacts with a cell with the motile strategy M. Table  \ref{tab:motprolifGrowth} shows that cells will have a higher proliferative potential when they interact with cells that have the motile strategy. Table \ref{tab:motprolifMot} shows that interacting with cells with the motile strategy will also increase the probability of motility of the interacting cell. These tables are used to guide the behaviour of the cells on a 500x500 lattice containing initially 5 cells of each type and two unequal sources of nutrients in two different locations. In each time step the interactions of a cell with its neighbours are used to adjust the probabilities of proliferation and motility. 

\begin{figure}
	\centering
		 \includegraphics[scale=0.8]{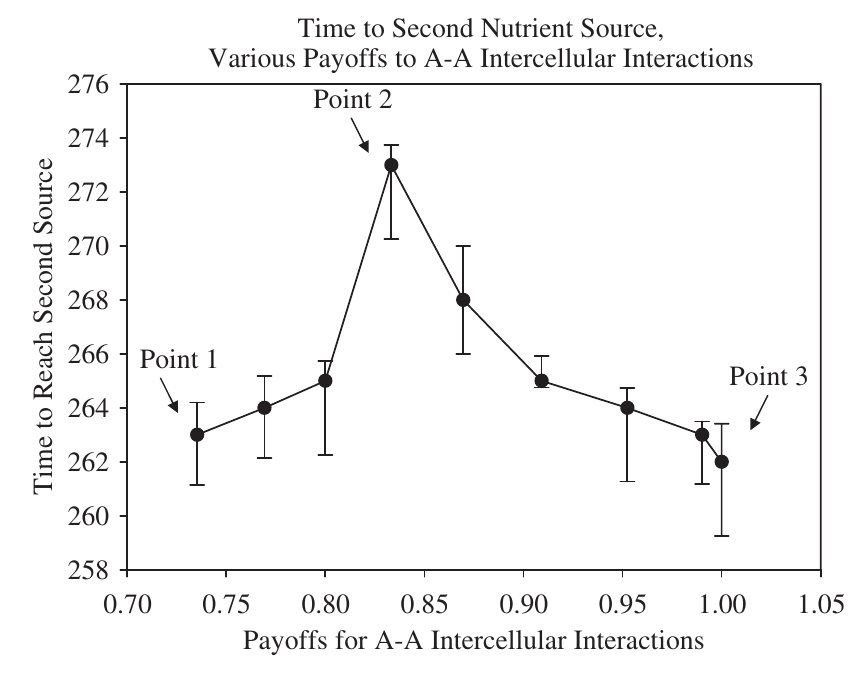}
	\caption{Plot of the time to reach a nutrient source (i.e. the inverse of the tumour's average velocity) versus various payoff values conferred to P-P intercellular interactions (where \textit{A} is the label for the proliferative phenotype in the paper). The error bars indicate the standard deviations from performing 10 Monte Carlo simulations for each value of payoffs for P-P interactions. (from \cite{mansury:2006} with permission).}
	\label{fig:deisboeck}
\end{figure}

\textbf {Results}.The authors used this model to study how varying the payoffs for A-A interactions affects a number of features of the tumour such as the speed (see figure \ref{fig:deisboeck}) at which it reaches the nutrient sources or the fractality of the resulting spatial patterns. 

\textbf {Remarks}. Although the model is interesting it is not as simple as other models reviewed in this chapter. A significant drawback is that the model is not evolutionary in the sense that it does not take into account the possibility of mutations introducing new phenotypes or tumour cells producing offspring different from their progenitors. This has been identified by the authors as something to be addressed in a future version of the model although they still claim the model to be based on EGT. Moreover, the authors do not take advantage of the tools provided by GT analysis to study the different steady state situations that could arise for different values of the payoff tables. This is probably due to the fact that the model is not a conventional GT model and the conventional game theoretical tools would not be easy to use in this context. Also, the model provides a view on the dynamics of the tumour growth that is rarely found in most other game theory analysis whose focus is on the study of equilibria.

\textbf{Model II}. The emergence of invasive phenotypes is influenced not only by its interaction with one phenotype or the other but by the complex interplay of several phenotypes which, in many cases has an indirect effect. Basanta et al. \cite{Basanta:2007} hypothesise a  number of scenarios in which three types of phenotypes interact in games with two and three players. They place these phenotypes in the context of an evolutionary non-spatial game theoretical model to test the hypothesis by Gatenby and colleagues that tumour invasion is promoted by the emergence of cells with a glycolytic metabolism \cite{Gatenby:2006}. The model represents  a glioma tumour populated by cells with enhanced proliferative capabilities, known as autonomous growth cells (AG), which can mutate into cells whose phenotype can make then follow either a more motile strategy (INV) or the glycolytic metabolism strategy (GLY).

\begin{table}[h]        
	\begin{center}
	\caption{Payoff table that represents the change in fitness of a tumour cell with a given phenotype interacting with another cell. Three different strategies are considered, those with higher replicative potential (AG), enhanced motility (INV) and glycolytic metabolism (GLY). The base payoff in a given interaction is equal to 1 and the cost of moving to another location with respect to the base payoff is $c$. The fitness cost of acidity is $n$ and $k$ is the fitness cost of having a less efficient glycolytic metabolism. The table should be read following the columns, thus the fitness change for an INV cell interacting with an AG would be $1-c$. }
	\begin{tabular}{|c|c|c|c|} \hline
	  &      AG & INV          & GLY     \\ \hline
	AG            &   $ \frac{1}{2}$ &       $1-c$ & $\frac{1}{2} +n-k$                         \\ \hline       
	INV             & 1& $1-\frac{c}{2}$ &   $1-k$       \\ \hline
	GLY & $\frac{1}{2}-n$  & $1-c$ & $\frac{1}{2} -k$ \\ \hline
	\end{tabular}  
	\label{tab:glymot}
	\end{center}
\end{table}

\textbf{Results}. Table \ref{tab:glymot} defines the interactions between the three phenotypes. The authors investigate a number of scenarios in which subgames based on two strategies are used to study how one phenotype could emerge in a tumour populated by cells that use a different strategy. The results show that cells with a higher replicative potential (AG) and invasive cells (INV) can coexist in a tumour as long as the fitness cost of motility is not too high. They also reveal that autonomous growth (AG) cells cannot coexist with glycolytic (GLY) cells. An interesting result of the model when all three strategies are considered simultaneously is that the appearance of the invasive phenotype is facilitated by the existence of glycolytic cells. Figure \ref{fig:3phenotypesPM} shows how the proportion of invasive cells (X axis) increases as the cost of having a glycolytic metabolism ($k$) decreases and the cost of living in an acid environment ($n$) increases. In other words, the success of the invasive phenotype depends on the same factors that determine the fate of the glycolytic phenotype. The model suggests that any therapy that could increase the fitness cost of tumour cells switching to a glycolytic metabolism or the susceptibility of normal cells to acid environments might decrease the probability of  the emergence of more invasive phenotypes.

\begin{figure}
	\centering
		 \includegraphics[scale=0.6]{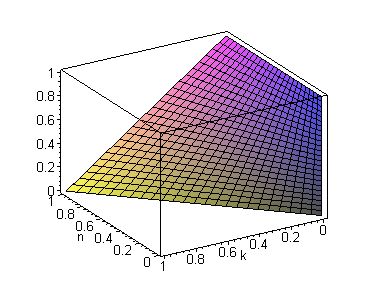}
	\caption{Proportion of invasive cells in a tumour with three phenotypes (autonomous growth, invasive and glycolytic). $k$ is the cost in terms of fitness of adopting the glycolytic metabolism whereas $n$ is the fitness cost of a normal cell when staying with a glycolytic cell.}
	\label{fig:3phenotypesPM}
\end{figure}


\section{Outlook}

The examples described in this chapter treat tumour populations as ecosystems of potentially co-operating and/or competing cells. In these ecosystems, the success of one phenotype depends on its interactions with other existing phenotypes. Such an approach has been shown to be a helpful and useful way to study cancer evolution. Most of the applications use GT to study the steady state of a population of tumour cells that follow different strategies dictated by their phenotypes, acquired as a result of genetic and epigenetic mutations. This can be very relevant to study the different ways in which a tumour population may evolve under different model parameters and assumptions of their interplay. Such studies could lead to cancer therapies that would alter the dynamics of cancer evolution towards benign tumours. One limitation of EGT/GT is that all relevant potential phenotypes/strategies have to be known a priori if a good understanding of cancer evolution is to be obtained. Knowing all the potential relevant phenotypes might be difficult and even if the phenotypes and their interactions are well known the EGT model might be too complicated to be analysed. Moreover, with more strategies/phenotypes the composition of the population may not converge to an equilibrium and the frequencies of the phenotypes could keep oscillating in a regular or chaotic fashion \cite{Sigmund:1999}. 

One more limitation of the GT models shown in this chapter is that they do not  study the dynamics in a tumour population (dynamics which may or may not lead to an equilibrium). GT models that make use of population biology have the potential to overcome this limitation and also ease the connection between a quantitative model and experimental data as the payoff tables used in more conventional EGT models tend to make the assumption that the fitness values are independent of space or time. One promising venue is to couple conventional GT with population dynamics which is also based on the assumption that successful strategies spread  \cite{Hoppensteadt:1982,Sigmund:1999}. This trend is shown in the work of Gatenby and Vincent \cite{gatenby:2003a} whose EGT model uses methods from population biology in order to study how the phenotypes of cells in a population evolve towards ESS. Gatenby and Vincent adopted a game theory approach influenced by population dynamics to study the influence of the tumour-host interface in colorectal carcinogenesis. The authors formulated an extended system of Lotka-Volterra equations to model the effect of nutrients and sensitivity to growth constraints in the proliferation of tumour cells. The cells in the tumour population are characterised by the number of substrate transporters (a higher number of them allow more nutrients in the cell) on the cell surface and by the cell response to normal growth constraints. Initially all the tumour cells are assumed to have normal values for both parameters but these values are allowed to change and the authors study their evolution. Their work demonstrates that normal cells in a multicellular organism occupy a ridge-shaped maximum of the fitness landscape that allows the heterogeneous coexistance of multiple cells. This fact makes them susceptible to mutations that are fitter and thus allow the somatic evolution that characterises cancer. The authors also conclude that any therapy that reduces the population density would be counterproductive since it would allow the more aggressive phenotypes to grow. Subsequent research uses the same formulation to study other stages of tumour progression like invasion and metastasis \cite{Vincent:2006,T.-L.-Vincent:2007}.

\begin{figure}
	\begin{center}
		\subfigure[]{\includegraphics[scale=0.6]{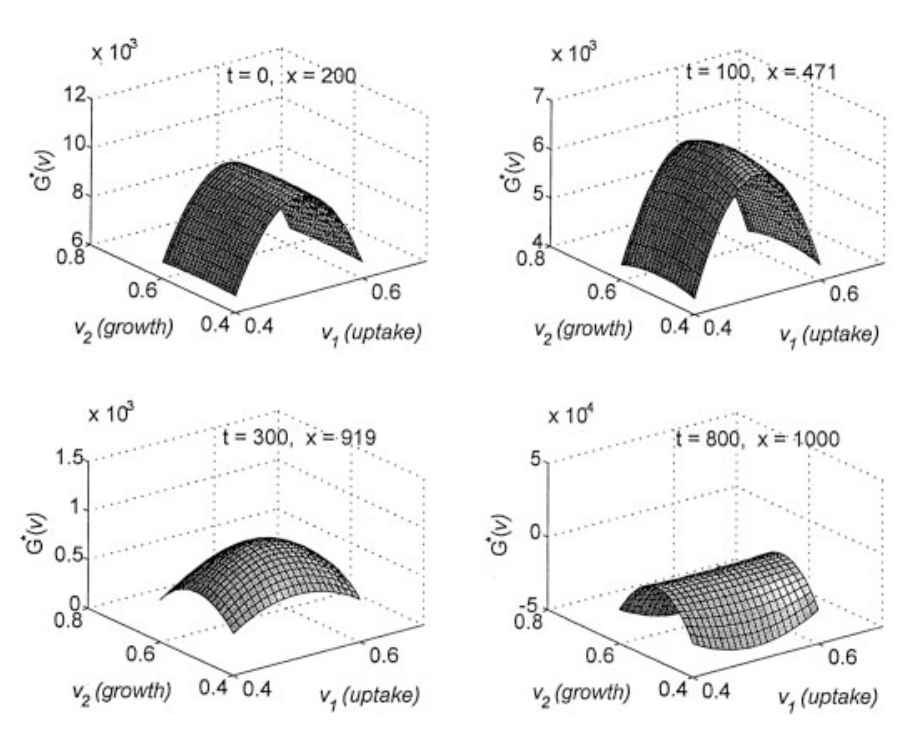}}\quad
		\subfigure[]{\includegraphics[scale=0.6]{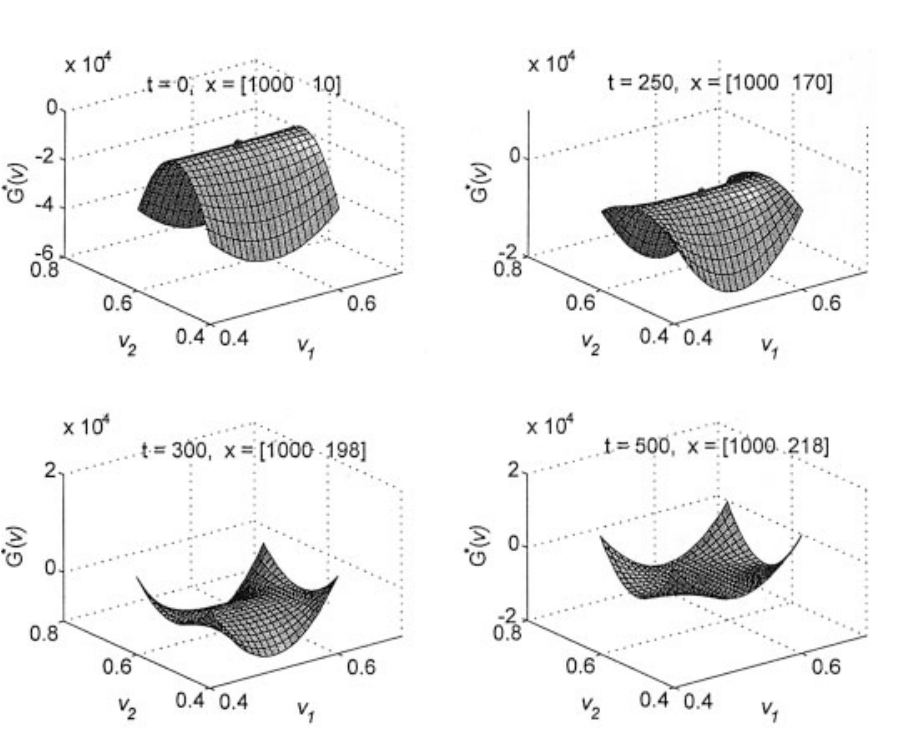}}
	\end{center}
	\caption{(a) Upon reaching equilibrium, normal cells sit on the ridge-shaped maximum in the in vivo adaptive landscape. This provides a tissue structure that allows coexistence of multiple distinct noncompeting populations, a necessary condition for the formation of multicellular organisms. (b) The presence of tumour cells deforms the tissue adaptive landscape over time so that normal cells arrive at a local minimum and are then readly sunject to invasion by the mutant phenotypes (from \cite{gatenby:2003a} with permission).}
	\label{fig:3phenotypesPM}
\end{figure}

GT is also a suitable tool to frame cooperative effects. Evolution of co-operation is a sub-field of GT pioneered by Robert Axelrod  \cite{Axelrod:1981}. Researchers in this field study the circumstances under which selfish agents will spontaneously co-operate. An example of how evolution of cooperation could be used to study cancer evolution was provided by Axelrod and collaborators in a recent paper \cite{Axelrod:2006} in which they show how tumour cells can cooperate sharing skills and capabilities (such as the production of angiogenic factors or paracrine growth factors needed to escape the homeostatic regulation of the tissue).


Evolutionary game theory has a relatively short history in the field of theoretical oncology but with an increasingly better understanding of the role of the microenvironment in tumour evolution \cite{Park:2000} and with the recent interest in studying cancer from ecological \cite{Crespi:2005,Merlo:2006} and phenotypic \cite{hanahan:2000} viewpoints, the role of this tool to understand the interactions between all the relevant agents within a tumour and their role driving cancer evolution will surely rise.

\section*{Acknowledgements}
We would like to acknowledge the help and suggestions from our colleagues at TU Dresden: Lutz Brusch, Haralambos Hatzikirou and Michael K\"{u}cken. The work in this paper was supported in part by funds from the EU Marie Curie Network  "Modeling, Mathematical Methods and Computer Simulation of Tumour Growth and Therapy" (EU-RTD-IST-2001-38923). We also acknowledge the support provided by the systems biology network HepatoSys of the German Ministry for Education and Research through grant 0313082C. Andreas Deutsch is  a member of the  DFG-Center for Regenerative Therapies Dresden - Cluster of Excellence - and gratefully acknowledges support by the Center.

\bibliography{chapterGT07}
\bibliographystyle{unsrt}
\end{document}